\documentclass[twocolumn]{aastex7}
\usepackage{hyperref}

\renewcommand{\phs}{\phantom{-}}
\usepackage{xcolor}

\begin{document}

\title{
Understanding the Origin and Dynamical Evolution of the Unique \\Open Star Cluster Berkeley 20 using FIRE Simulations}

\correspondingauthor{Alessa I.~Wiggins}

\author[orcid=0009-0008-0081-764X]{Alessa I.~Wiggins}
\affiliation{Department of Physics and Astronomy, Texas Christian University, TCU Box 298840 
    Fort Worth, TX 76129, USA }
\email[show]{a.ibrahim@tcu.edu}

\author[orcid=0000-0001-6086-9873]{Jamie R.~Quinn}
\affiliation{Department of Physics, University of California, Merced, 5200 North Lake Road, Merced, CA 95343, USA}
\email{jquinn4@ucmerced.edu}

\author[0000-0001-5636-3108]{Micah Oeur}
\affiliation{Department of Physics, University of California, Merced,
5200 North Lake Road, Merced, CA 95343, USA}
\email{moeur@ucmerced.edu}

\author[orcid=0000-0001-6761-9359]{Sarah R.~Loebman}
\affiliation{Department of Physics, University of California, Merced, 5200 North Lake Road, Merced, CA 95343, USA}
\email{sloebman@ucmerced.edu}

\author[orcid=0000-0002-0740-8346]{Peter M.~Frinchaboy}
\affiliation{Department of Physics and Astronomy, Texas Christian University, TCU Box 298840 
Fort Worth, TX 76129, USA }
\email{p.frinchaboy@tcu.edu}

\author[orcid=0000-0003-2594-8052]{Kathryne J.~Daniel}
\affiliation{Department of Astronomy \& Steward Observatory, University of Arizona,
Tucson, AZ 85721, USA}
\email{kjdaniel@arizona.edu}
 
\author{Fiona McCluskey}
\affiliation{Department of Physics \& Astronomy, University of California, Davis, CA 95616, USA}
\email{fmccluskey@ucdavis.edu}

\author[0000-0003-2602-4302]{Jonah M.~Otto}
\affiliation{Department of Physics and Astronomy, Texas Christian University, TCU Box 298840 
Fort Worth, TX 76129, USA }
\email{j.otto@tcu.edu}

\author[orcid=0000-0002-5267-3034]{Hannah R. Woodward}
\affiliation{Department of Astronomy, University of Wisconsin-Madison, 
475 N. Charter Street, Madison, WI 53706-1507 USA}
\email{hrwoodward@wisc.edu}

\author[orcid=0000-0003-2676-8344]{Elena D'Onghia}
\affiliation{Department of Astronomy, University of Wisconsin-Madison, 
475 N. Charter Street, Madison, WI 53706-1507 USA}
\email{edonghia@astro.wisc.edu}

\author[0000-0003-0603-8942]{Andrew Wetzel}
\affiliation{Department of Physics \& Astronomy, University of California,
Davis, CA 95616, USA}
\email{arwetzel@gmail.com}

\author[0009-0007-3431-4269]{Hanna Parul}
\affiliation{LIRA, Observatoire de Paris, Université PSL, Sorbonne Université, Université Paris Cité, \\CY Cergy Paris Université, CNRS, F-92190 Meudon, France}
\email{hparul@crimson.ua.edu}

\author[0000-0002-7707-1996]{Binod Bhattarai}
\affiliation{Department of Physics \& Astronomy, University of Alabama, Box 870324, Tuscaloosa, AL 35487-0324, USA}
\email{bbhattarai2@ua.edu}

\author[0009-0002-0600-7935]{Maximilian Cozzi}
\affiliation{Department of Physics, University of California, Merced, 5200 North Lake Road, Merced, CA 95343, USA}
\email{mcozzi@ucmerced.edu}

\begin{abstract}
Open clusters (OCs) act as key probes that can be leveraged to constrain the formation and evolution of the Milky Way (MW)'s disk, as each has a unique chemical fingerprint and well-constrained age. 
Significant Galactic dynamic interactions can leave imprints on the orbital properties of OCs, allowing us to use the present day properties of long-lived OCs to reconstruct the MW’s dynamic history.
To explore these changes, we identify OC analogs in FIRE-2 simulations of MW-mass galaxies.   
For this work, we focus on one particular FIRE-2 OC, which we identify as an analog to the old, subsolar, distant, and high Galactic latitude MW OC, Berkeley 20.
Our simulated OC resides $\sim$6 kpc from the galactic center and ultimately reaches a height $|Z_{\mathrm{max}}|>2$ kpc from the galactic disk, similar to Berkeley 20. We trace the simulated cluster's orbital and environmental history, identifying key perturbative episodes, including: (1) an interaction with a gas overdensity in a spiral arm that prompts an outward migration event and (2) a substantial interaction with a {Sagittarius Dwarf Spheroidal} Galaxy-mass satellite that causes significant orbital modification. 
Our simulated OC shows significant resilience to disruption during both its outward migration and the satellite-driven heating event that causes subsequent inward migration.
Ultimately, we find these two key processes --- migration and satellite heating --- are essential to include when assessing OC orbital dynamics in the era of Gaia. 

\end{abstract}


\section{Introduction} 
Open clusters (OCs) are vital tracers of local star formation in nearby galaxies \citep[e.g.,][]{Maschmann2024} and provide critical constraints on the structure and evolution of the Milky Way (MW)'s disk \citep[e.g.,][]{Donor2020, Spina2021}.
Large-scale astrometric and spectroscopic surveys like \textit{Gaia} \citep{Gaia2016} and APOGEE \citep{APOGEE2017} have enabled increasingly complete OC samples with well-determined positions, kinematics, ages, and chemical abundances \citep[e.g.,][]{Cantat-Gaudin2020, Myers2022, Hunt2024}. 
This wealth of precise data has transformed our ability to identify and study outlier clusters with atypical chemistry, orbits, or ages relative to their position in the disk \citep[e.g.,][]{Cantat-Gaudin2016}. 
Such rare systems offer a unique perspective into the MW’s dynamic history. One such cluster is Berkeley 20 (Be20), a long-known but increasingly compelling outlier.

Originally identified as an unusual system by \citet{MacMinn1994}, Be20 is an old \citep[4.79 Gyr;][]{Cantat-Gaudin2020}, distant \citep[Galactocentric radius of 16.32 kpc;][]{Cantat-Gaudin2020} OC with subsolar metallicity ([Fe/H]~$\approx -0.42$ dex; \citealt{Cantat-Gaudin2016, Myers2022}), and an extreme vertical position, currently located 2.61 kpc below the Galactic midplane \citep{Cantat-Gaudin2020}. 
Orbit integrations using \texttt{Gala} \citep[version 1.9.1, with MW potential 2022;][]{Price-Whelan2024} calculate the mean angular momentum of the orbit and yield a guiding radius\footnote{Guiding radius is defined here as the radius of circular orbit with the same angular momentum.} of 15.16 kpc. These results also indicate that the cluster is currently at its maximum vertical excursion \citep[][]{Myers2022, Otto2025}.

Kinematic measurements further support Be20’s status as a dynamic outlier. 
\citet{Friel2002} identified six members with radial velocities of $+70 \pm 13$ km s$^{-1}$, consistent with the values reported by \citet{Frinchaboy2006}, and more recently, \citet{Myers2022} measured an RV of $+76.6 \pm 0.2$ km s$^{-1}$ using APOGEE data. 
Early orbit reconstructions by \citet{Wu2009} found Be20 to have a highly eccentric, vertically extended orbit, characteristics they interpreted as evidence for thick disk membership. \citet{VandePutte2010} proposed an alternative scenario, suggesting the orbit might point to an extragalactic origin. 
However, more precise {\em Gaia} proper motions have enabled updated orbital integrations, which reveal a significantly lower eccentricity of 0.087 and relatively regular orbital parameters with an average vertical oscillation period of 374 Myr and a radial period of 323 Myr \citep[][]{Myers2022,Otto2025}. 
In fact, despite its large vertical displacement, Be20 appears to follow a nearly circular orbit. 
High resolution spectroscopic analysis by \citet{Cantat-Gaudin2016} suggests that the cluster may be a perturbed thin disk object or a genuine thick disk system influenced by accretion-driven dynamics. 
They emphasize that Be20’s origin remains deeply uncertain and merits substantive investigation.

Notably, in the era of {\em Gaia}, the MW’s disk is increasingly understood as a dynamically evolving structure shaped by both internal and external interactions.
Stars, and presumably gravitationally bound
OCs, can be redistributed across the disk through resonant interactions with internal structures such as a bar or spiral arms.
A process, colloquially known as radial migration \citep[also known as and henceforth referred to as cold torquing,][]{daniel&wyse2018}, causes stars and clusters to have large changes in their mean orbital radii without significant changes in their orbital eccentricity \citep{SellwoodBinney2002}.
Cold torquing specifically occurs at the corotation radius of transient spirals. 
However, other resonances from both spirals and bars \citep[e.g.,][]{BarbanisWoltjer1967, CarlbergSellwood1985} and overlapping resonances \citep{Minchev2011,Daniel19} change orbital radii but also kinematically heat.
Satellite interactions and mergers are also known to vertically heat orbits \citep[e.g.,][]{QuinnHernquistFullagar1993, Bird2012}; additionally, scattering by GMCs \citep[e.g.,][]{SpitzerSchwarzschild1951} can convert in-plane motions to vertical excursions \citep[e.g,][]{Lacey1984}. 
The impact of each of these mechanisms has been explored both in isolation, especially in controlled experiments and idealized models \citep[e.g.,][]{Roskar2012,Daniel15,Quillen18}, and in cosmological simulations, which have examined the impact of spiral structure and bars on vertical heating \citep[e.g.,][]{grand2016}.
However, cosmological hydrodynamic simulations have yet to provide a comprehensive, orbit-level connection between large scale structures, like spiral features, and individual stellar and OC trajectories. 

Among the most significant of the external perturbations for the MW is the {Sagittarius Dwarf Spheroidal Galaxy (Sgr dSph)}, which has likely completed several pericentric passages on a near polar orbit over the last 6--8 Gyr 
\citep{Purcell2011, Laporte2018, Antoja2020}. 
These events have been shown to trigger warps, vertical oscillations, spiral structure, and broad-scale radial migration. 
Simulations by \citet{Purcell2011} and \citet{Laporte2018} demonstrate that repeated {Sgr dSph} passages can excite bending and breathing modes, drive phase-space spirals, and redistribute stars throughout the disk \citep{Antoja2018}. 
Recent work by \citet{Carr2022} using a collisionless $N$-body simulation highlights how such interactions displace stars from near-circular orbits, altering guiding centers and eccentricities in a coherent, quadrupole pattern. 

While these studies underscore the disk-wide dynamic consequences of satellite interactions, the long-term survival and migratory history of individual OCs in this cosmological context remains unexplored.
In this work, we use high-resolution hydrodynamic cosmological simulations that self-consistently form OCs and follow their orbital evolution over time. 
In particular, we follow one such cluster that migrates outward, directly experiences satellite-driven perturbations, then migrates inward and ultimately survives to the present day. 
This approach offers a new window into how complex interactions shape the orbital evolution of clusters like Be20.

\section{Methods} 

\begin{deluxetable}{lcc}[h!]
\tabletypesize{\footnotesize}
\tablecaption{Comparison of Simulated Cluster (BOB) and Observed Cluster Berkeley 20 (Be20)}
\label{Table:properties}
\tablehead{\colhead{Property} & \colhead{BOB (simulated)} & \colhead{Be20 (observed)}}
\startdata
Age (Gyr)                           & $\phs5.13$                   & $\phs4.79$\tablenotemark{a}\\
Mass ($M_\odot$)                    & $\phs2.6\times10^4$\tablenotemark{b}                   & $\phs1\times10^3$\tablenotemark{c}\\
Eccentricity                        & $\phs0.18$                   & $\phs0.09$\tablenotemark{d} \\
$\overline{\mathrm{[Fe/H]}}$ (dex)  & $-0.27$                      & $-0.42$\tablenotemark{d,g} \\
$Z_{\mathrm{max}}$ (kpc)            & $-2.75$                   & $-2.61$\tablenotemark{a} \\
$Z_{\mathrm{present}}$ (kpc)        & $\phs0.43$                   & $-2.61$\tablenotemark{e} \\
$R_{\mathrm{present}}$ (kpc)        & $\phs5.80$                   & $16.32$\tablenotemark{a} \\
$R_{\mathrm{guiding}}$ (kpc)        & $\phs6.27$                   & $15.15$\tablenotemark{d} \\
$R_{\mathrm{max}}$ (kpc)            & $11.49$\tablenotemark{f}     & \nodata \\
$R_{\mathrm{birth}}$ (kpc)          & $\phs5.80$                  & \nodata \\
Early Eccentricity                  & $\phs0.27$\tablenotemark{h}                    & \nodata \\
$\sigma_{1D}$ (km s$^{-1}$)              & $\phs1.34$                   & \nodata \\
\enddata
\tablenotetext{a}{\citet{Cantat-Gaudin2020}}
\vspace{-2 mm}
\tablenotetext{b}{Minimum resolvable cluster mass \citep[see][]{Bhattarai2024}.}
\vspace{-2 mm}
\tablenotetext{c}{\citet{Gloria2011}.} 
\vspace{-2 mm}
\tablenotetext{d}{Computed in \citet{Otto2025} from \citet{Cantat-Gaudin2020} using \citet{Price-Whelan2024}}
\vspace{-2 mm}
\tablenotetext{e}{\citet{Cantat-Gaudin2016}}
\vspace{-2 mm}
\tablenotetext{f}{\citet{Myers2022}}
\vspace{-2 mm}
\tablenotetext{g}{Effective stellar disk edge is 13.5 kpc \citep{bellardini}.
\vspace{-2 mm}
\tablenotetext{h}{Calculated near cluster formation}}

\end{deluxetable}

We analyze an OC drawn from \texttt{m12f}, a cosmological zoom-in simulation of a MW–mass galaxy from the \textit{Latte} suite \citep{Wetzel2016} within the FIRE-2 project \citep{Hopkins2018}. 
The simulation adopts a $\Lambda$CDM cosmology and outputs 600 snapshots from redshift $z = 99$ to $z = 0$ at $\sim$25 Myr intervals, enabling detailed orbital tracking. 
Notably, \texttt{m12f} undergoes a late-time interaction with a near Sgr dSph-mass satellite that triggers both a transient bar \citep{Ansar2025} and transient spiral structure \citep{quinn2025}, providing a rich dynamic environment for studying cluster kinematic evolution.

The satellite responsible for the interactions we study reaches its first perigalactic passage at $t = 10.79$ Gyr into the simulation ($z = 0.26$), passing within $\sim$36 kpc of the host center with a total mass of $6.0 \times 10^{10}\,M_\odot$ and stellar mass of $2.6 \times 10^9\,M_\odot$; the satellite merges at $t = 12.36$ Gyr ($z = 0.11$).
{This satellite is similar in mass to Sgr dSph at first passage \citep[e.g.,][]{Laporte2018} and follows an orbit with a comparable pericentric distance but a less polar inclination ($\mathbf{\sim 10^\circ}$ below the midplane) than Sgr dSph. }
Despite this, the satellite serves as a reasonable first pass analog for studying the potential dynamic impact of Sgr dSph-like satellites on outer-disk OCs such as Be20.

In \texttt{m12f}, star particles and gas cells have typical initial masses of $7070 M_\odot$, with minimum gravitational softenings of 4 and 1 pc, respectively. 
Star formation occurs in cold, dense, molecular, self-gravitating gas, with explicit modeling of feedback from supernovae, stellar winds, radiation pressure, and photoionization. 
At \textit{Latte}'s resolution, individual star particles represent mono-age, mono-abundance populations \citep[see][for more details]{Weztel2023}, analogous to segments of real star clusters.

We identify OCs using the method described in \citet{Bhattarai2024}, which applies a friends-of-friends algorithm with a 4 pc linking length.
We require at least five star particles per cluster for robust statistics; at redshift $z = 0$, typical particle masses in old OCs are $\sim5000\,M_\odot$, setting a minimum resolved cluster mass of $\sim10^{4.4}\,M_\odot$. 
The resulting systems are compact ($R_{1/2} \sim 3$ pc), kinematically cold ($\sigma_{1D} \sim 4$ km s$^{-1}$), and nearly coeval ($\sigma_{\rm age} \sim 0.25$ Myr).

\citet{WigginsPHD} presents a catalog of 128 OCs older than 1 Gyr found in three \textit{Latte} galaxies at present day, including 28 from \texttt{m12f}. 
From this sample, we identify one cluster with age, metallicity, and orbital characteristics similar to Be20 (see Table~\ref{Table:properties}). 
We refer to it as “BOB” (for \underline{B}eaten \underline{O}pen cluster \underline{B}erkeley 20 analog) and use it to explore a plausible evolutionary pathway for Be20's present-day properties.
To consider the impact of the local environment on BOB's orbital evolution, at each timestep, we compute {both the stellar and gas densities in two regions: 
(1) A 100 pc radius sphere centered on BOB, which defines the local density ($\rho_{\mathrm{local}}$), and
(2) an annular shell at the same Galactocentric radius with a thickness and height of 200 pc that captures the typical global density conditions ($\rho^R_{\mathrm{annulus}}$).
Their ratio, $\rho_{\mathrm{local}} / \rho^R_{\mathrm{annulus}}$, traces the cluster’s relative environment over time.}

\begin{figure*}
\includegraphics[width=.95\textwidth]{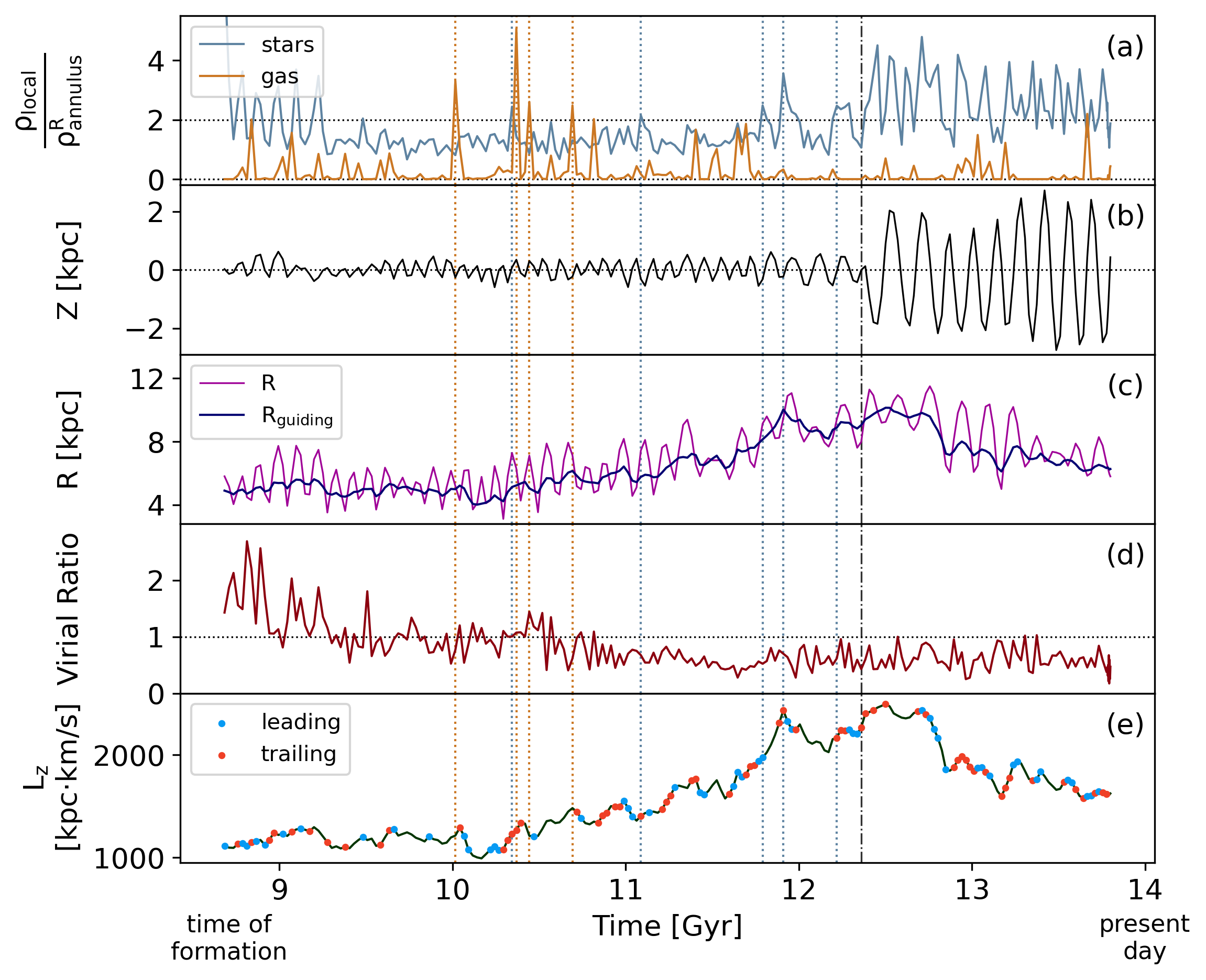}
\caption{\textbf{BOB undergoes major orbital changes}. 
Panel a shows {{the ratio of the local to annular stellar (blue) and median gas (dark orange) densities over time. Blue/dark orange vertical dotted lines indicate when BOB encounters a significant stellar/gaseous overdensity. We identify stellar density peaks after the disk settles at $\sim$9.25 Gyr \citep[][]{Yu2021} and before the merger event.} {The black, right-most dash-dotted} line at $\sim$12.4~Gyr indicates the merger event with a $\sim$Sgr dSph-mass satellite, after which {BOB's vertical amplitude increases} dramatically.}
Panel b tracks BOB's height above/below the midplane, while panel c shows its instantaneous and guiding radius over time.
Panel d shows BOB's virial ratio, which stays $\lesssim1$ after 10 Gyr, indicating the cluster remains bound.
Panel e shows the cluster's angular momentum, $L_z$, with blue/red points indicating BOB's location on the leading/trailing edge of a dominant spiral arm, respectively, otherwise BOB is in an inter-arm region.}
\label{figure:4panel}  
\end{figure*}

\section{Results}
\label{sec:results}

\begin{figure*}
\includegraphics[width=.95\textwidth]{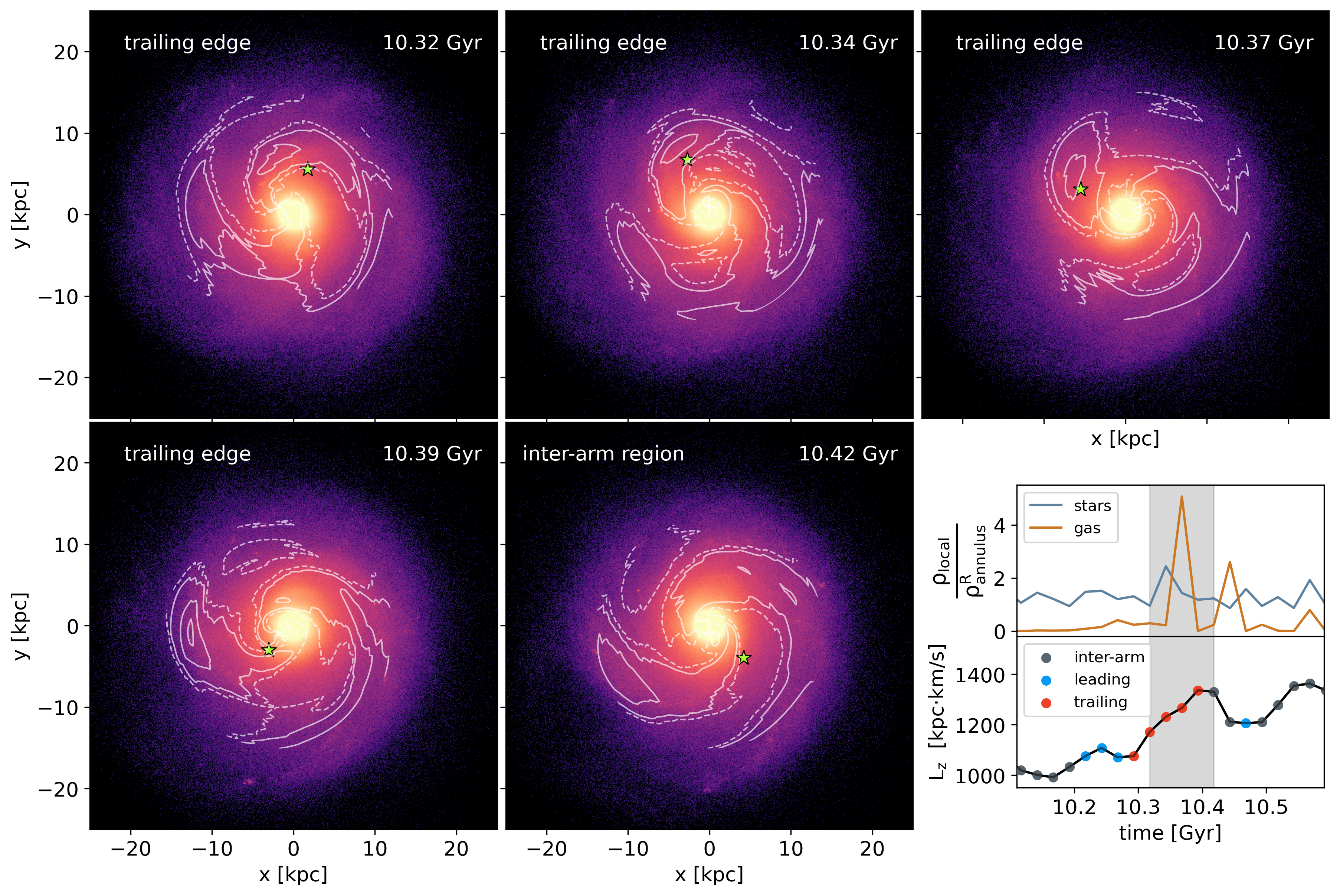}
\caption{\textbf{BOB experiences an increase in $L_z$ while on the trailing edge of a dominant spiral arm.}
{The stellar mass distribution is shown with solid/dashed contours \citep{quinn2025} indicating the 25\%, 15\%, 2.5\% over/under densities.
BOB is indicated by the green star.
The bottom right panel shows the {ratio of the local to annular stellar and median gas density} (top) and $L_z$ (bottom), with red/blue points marking BOB's position on the trailing/leading edge of a dominant spiral arm. 
The shaded region highlights the time window shown in the five top-down panels (10.32–10.42 Gyr).
BOB increases in $L_z$ while it interacts with the trailing edge of the spiral arm, consistent with expectations from cold torquing \citep{daniel&wyse2018} and corresponding to a quick rise in  $R_{guiding}$ without a large change in vertical amplitude. {An animation of BOB’s dynamic evolution from 8.68 to 13.78 Gyr is available in the online Journal. The animation includes two panels. Left: Top-down view of BOB (green star) overlaid on the stellar mass distribution in the m12f simulation with solid/dashed contours showing the stellar over/under densities. Right: time-series of major orbital changes in BOB (as in Figure~\ref{figure:4panel}).}}}
\label{figure:contour_plots}  
\end{figure*}

\begin{figure*}[ht!]
    \centering
    \includegraphics[width=0.95\textwidth]{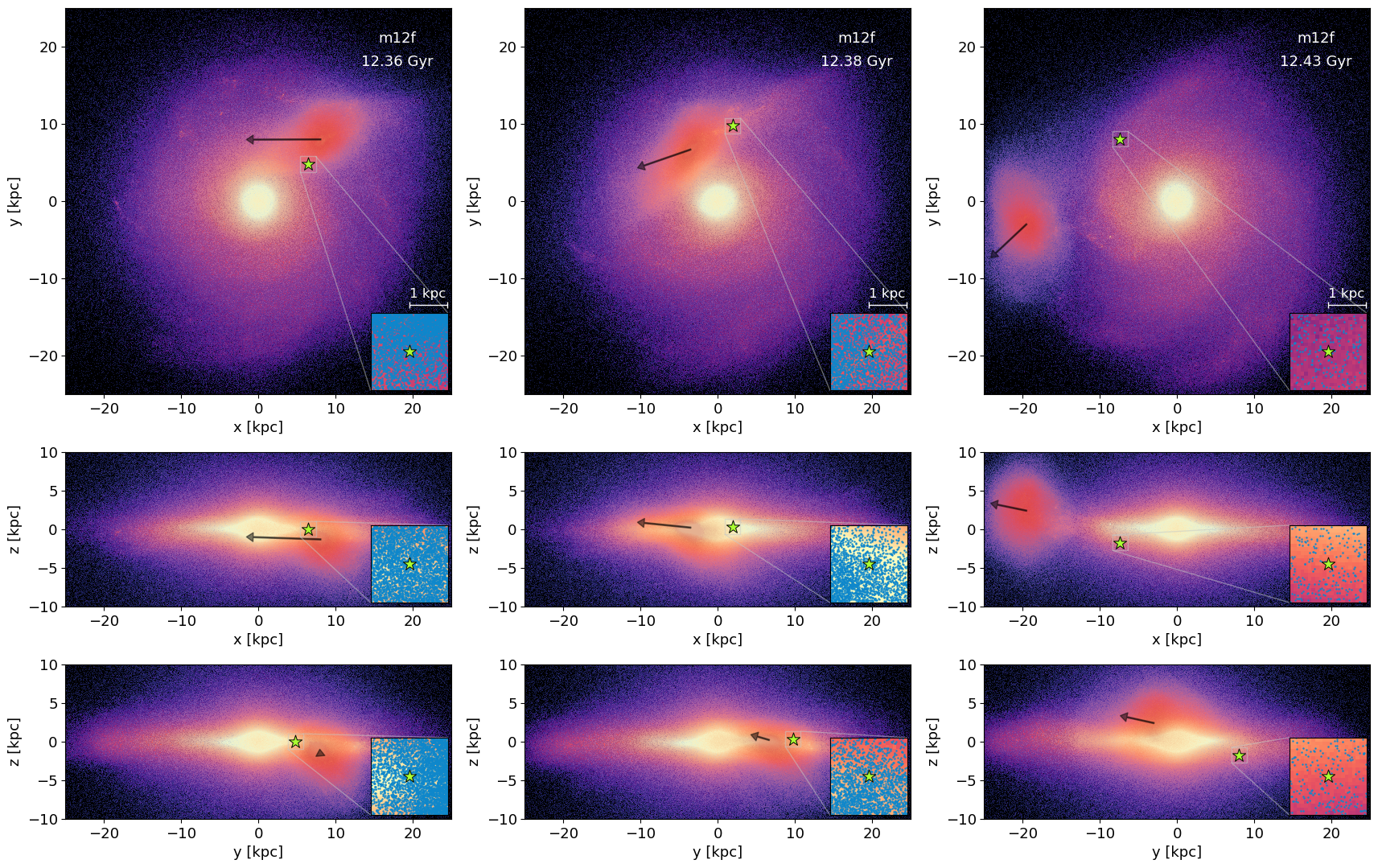}   
    \caption{\textbf{BOB's vertical height changes significantly in response to the satellite's passage}.
    Top-down and side-on visualization of insitu and exsitu stars in \texttt{m12f} during and after the satellite interaction. BOB is marked in green in each panel, with a 4~kpc$^2$ zoom centered on BOB. In-situ stars are shown in the background, and ex-situ stars are highlighted in blue. Arrows indicate the satellite’s center of velocity vectors, originating from its center of mass, computed via an iterative zoom-in method.
    {\em Left:} Satellite appears on the lower right edge-on. The region surrounding BOB is well populated with satellite stars, and BOB is near the mid-plane of the disk. 
    {\em Middle:} Satellite passes through the disk, and the bulk of its stars pass near BOB. 
    {\em Right:} Satellite is at the outskirts of the disk, above the mid-plane. There are few exsitu stars near BOB, but the impact of the satellite interaction can be seen by the drastic change in $Z$. 
    }
    \label{figure:path}
\end{figure*}

\begin{figure}[ht!]
    \centering
    \includegraphics[width=0.49\textwidth]{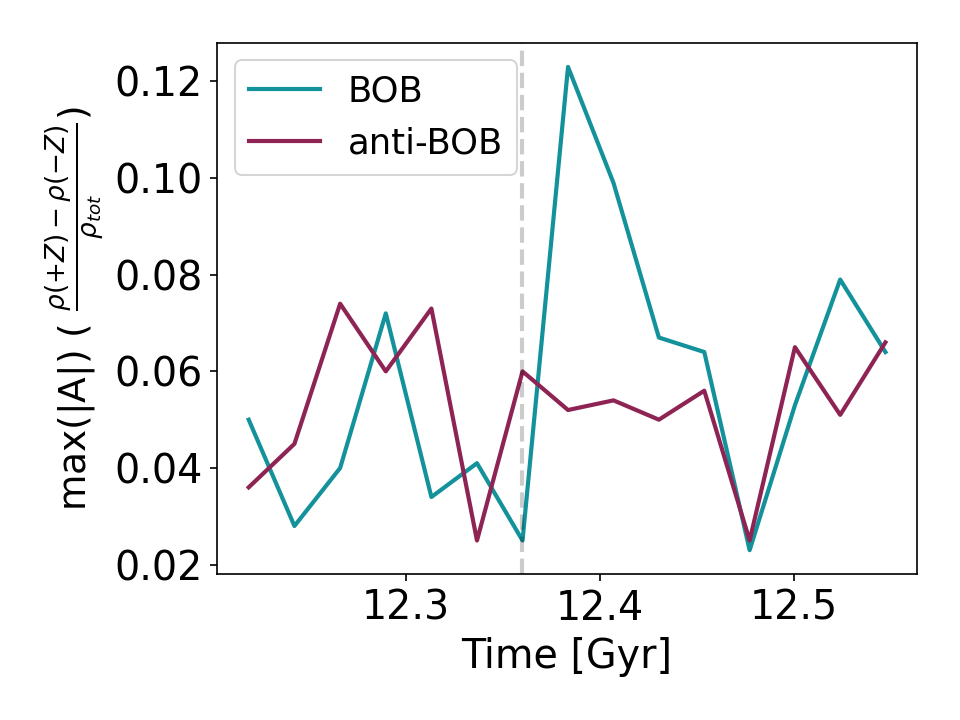}
    \caption{
    \textbf{The satellite's passage induces significant asymmetry in the stellar mass distribution, $\mathbf{\rho}$(Z), about the midplane.} We measure the asymmetry within a cylindrical volume centered on BOB with a radius of 1.4 kpc and height of $\pm3$ kpc. {The stellar disk near BOB is twice as asymmetric as anti-BOB, which corresponds to the same cylindrical region but on the opposite side of the disk azimuthally immediately following the satellite passage (marked by the dashed gray line). This spike is evidence of a strong, localized vertical mass redistribution in the disk surrounding BOB, a direct response of the satellite encounter.}}
    \label{figure:asym}
\end{figure}

We find that BOB, the simulated Be20 analog, undergoes significant and varied orbital modifications throughout its multi-Gyr dynamic lifetime due to both internal and external drivers.

\subsection{Internally driven dynamic evolution}

First, we analyze in detail the cluster's evolution due to its local environment.
{Figure~\ref{figure:4panel}a shows the $\rho_{\mathrm{local}} / \rho^R_{\mathrm{annulus}}$ ratio in both stars and gas over time.}
{We indicate moments when BOB passes through an overdense region of stars/gas by the blue/dark orange vertical dotted lines.}
{We note that BOB's large vertical height following the merger event (vertical black dashed-dotted line) results in a significant increase in stellar $\rho_{\mathrm{local}} / \rho^R_{\mathrm{annulus}}$, and we do not highlight peaks over this period.}
{The highlighted instances of BOB passing through a gas or stellar overdensity generally corresponds to an interaction with a spiral arm and a subsequent change in angular momentum (Figure~\ref{figure:4panel}e), which is consistent with theoretical expectations \citep[][]{Modak2025, daniel&wyse2018}}.

BOB passes through a particularly significant {stellar and gaseous} overdensity in a spiral arm at $\sim$10.4~Gyr. 
Following this event, BOB begins to move outward from its birth location of $R$ =5.80 kpc, as seen in Figure \ref{figure:4panel}c. 
This outward movement persists for $\sim$2 Gyr, with BOB ultimately reaching $R\sim$10 kpc at 12.4 Gyr.

Interestingly, BOB's vertical displacement is largely unchanged over the course of its outward journey, remaining close to the midplane during this period of time (Figure \ref{figure:4panel}b).
This stands in contrast to the general trend in FIRE, which shows that the velocity dispersion of stars increases rather significantly with age due to post-formation scattering interactions \citep[e.g., asymmetric drift;][]{Yu2022}. For example, \citet{McCluskey2024} shows that, on average, the vertical velocity dispersion of stars increases by $\sim$7.5 km/s/Gyr near the end of their simulations.
It is possible that BOB's outward movement helps facilitate its continued kinematic coldness, as the density of scatterers (e.g., Giant Molecular Clouds (GMCs)) tends to decrease with increasing galactocentric radius. 
We assess the boundedness of BOB over time (virial ratio = $2E_{\text{kin}}/|E_{\text{pot}}|$) in Figure \ref{figure:4panel}d. Similar to what has been seen previously \citep[e.g.,][]{Kim2018}, BOB is slightly less bound at formation and becomes more bound over time. Interestingly, BOB remains highly bound (virial ratio $\lesssim1$) during its outward movement throughout the disk.

Next, we consider BOB's angular momentum ($L_{z}$) in Figure \ref{figure:4panel}e. 
We observe small-scale increases/decreases in $L_{z}$ throughout its lifetime, which coincides with BOB's instantaneous location relative to the trailing/leading edges of the spiral structure, respectively. 
{The changes in $L_z$ are also highlighted by the vertical dotted lines that indicate when BOB interacts with a stellar/gaseous overdensity.}
After BOB interacts with the overdense region of {both stars and gas} at $\sim$10.4 Gyr, we also identify a steady increase in $L_{z}$. This, in combination with the small-scale oscillations in $L_{z}$, is indicative of cold torquing \citep{roskar2008, daniel&wyse2018}.  

To assess this more deeply, in the bottom right panel of Figure \ref{figure:contour_plots}, we investigate the initial increase in $L_z$. In the other panels, we show BOB’s top-down location in the stellar disk at five moments in time corresponding to when the cluster passes through the {stars and} cold dense gas. The solid/dashed contours from \cite{quinn2025} show the over/under densities in the stellar mass distribution, highlighting where BOB is spatially located relative to the dominant spiral arms. We consider BOB to be on the trailing edge of a spiral arm if it is behind the azimuth of maximum overdensity and on the leading edge if it is ahead of the azimuth of maximum overdensity. The spirals are present in all stellar age populations and persist on Gyr timescales, suggesting transient density wave-driven structures that are classically associated with cold torquing. BOB is not scattered over this period of time, rather BOB’s $L_z$ steadily increases while it is on the trailing edge of the dominant spiral arm. 
In the last moment in time, shown in the bottom middle panel of Figure \ref{figure:contour_plots}, BOB is in an inter-arm region after it has gained $L_z$ from interacting with the spiral arm.

While we highlight this particular instance, this is a recurrent phenomenon that can be seen between $\sim 10.4-12.4$~Gyr, often coinciding with the vertical dotted lines.
During this $\sim$2 Gyr period, interactions with spiral arms gradually increase BOB’s angular momentum ($L_z$) and guiding radius ($R_{guiding}$=$L_z / v_{\text{circ}}$) without a significant increase in vertical height, displaying a clear example of cold torquing. 

\subsection{Dynamical evolution influenced by external perturbation}

At $\sim$12.4 Gyr, BOB undergoes a major interaction (marked by the right-hand vertical line in Figure~\ref{figure:4panel}) that substantially alters its orbital parameters, dramatically increasing its maximum vertical excursions. 
As shown in Figure~\ref{figure:4panel}b, BOB's vertical height increases significantly, reaching a maximum height of 2.75 kpc below the midplane, similar to Be20 (see Table \ref{Table:properties}). This increase in the extent of BOB's height and concurrent inward migration are in agreement with theoretical expectations regarding the effect of satellite interactions on stellar kinematics. Analytic work and simulations find that satellites kinematically heat stellar orbits \citep[e.g.,][]{Velazquez1999, D'Onghia2016, Carr2022} and increase radial redistribution \citep[e.g.,][]{Quillen2009}, with stars in the outer disk with low velocity dispersion being the most prone to satellite-driven inward migration \citep{Bird2012}. 
Remarkably, BOB remains gravitationally bound (Figure \ref{figure:4panel}d) despite this major interaction (right vertical line) with the satellite galaxy.

The passage of the satellite through the disk and its proximity to BOB over time is shown in Figure \ref{figure:path}\footnote{See also animation linked in online version of the text.}. 
Note that while the exsitu stars (satellite stars shown in blue in the inset panels) pass near BOB, an insubstantial fraction are left behind. As the satellite begins passing through the disk, BOB lies squarely in the midplane just above the satellite (see the bottom left panel of Figure \ref{figure:path}).
At 12.38 Gyr, the satellite continues its passage through the disk and just passes by BOB. 
By 12.43 Gyr, two things have happened: (1) BOB is offset downward substantially, and (2) a spiral arm rises up in strength (in response to the satellite that just passed through). This arm is comprised of insitu stars and has a large amount of power coincident with the X-Y location of BOB in the disk. 
BOB effectively surfs this structure inward after this event.

Why does BOB’s maximum vertical excursion increase so dramatically following the satellite’s passage?
Is this behavior unique to BOB, or is it common among nearby disk stars?
Prior to the satellite–disk interaction, BOB resides in a mildly warped region of the outer disk (see Figure~\ref{figure:asym}, which quantifies the vertical asymmetry in the stellar mass density about the midplane, teal line).
Despite this initial warp, BOB’s maximum vertical displacement remains modest ($|Z_{max}| = 0.62$ kpc).
Following the satellite's passage; however, this region becomes \textit{significantly} more warped, by roughly a factor of two.
This sharp increase in vertical asymmetry occurs immediately after the satellite encounter (marked by the gray dashed line in Figure~\ref{figure:asym}), indicating a substantial imbalance in stellar mass about the midplane.
This enhanced asymmetry is not mirrored on the opposite side of the disk (shown in magenta), highlighting that the warping is highly localized to the region surrounding BOB,
and reflects a coherent distortion of the outer disk structure. For reference, we consider two regions immediately adjacent to BOB, finding asymmetry strengths lower than those of the region surrounding BOB but higher than those of the region on the opposite side of the disk.

As BOB migrates inward over time, it retains this vertical energy. Its present-day $|Z_{\rm max}|$ encodes a dynamic memory of the past perturbation.
We suggest that Be20’s large vertical excursion is a natural consequence of a localized, satellite-induced warp, such as might arise from a past passage of Sgr dSph.

\section{Conclusions} 

BOB presents a rare and richly informative case study of old OC dynamic evolution in a cosmological context. 
By tracing its migration history, kinematic evolution, and response to external perturbations, we gain new insight into the diverse physical processes that shape Be20. 
We find:

\begin{enumerate}
    \item {\bf Outward Migration Without Heating:}  
    BOB migrates outward from R$\sim$6 to 10 kpc over $\sim$2 Gyr following a passage through a region of dense gas, corresponding to an interaction with a spiral arm. Despite this substantial radial shift, the cluster's vertical amplitude remains remarkably stable. 
    This behavior is consistent with cold torquing, where the cluster gains angular momentum from interactions with transient, non-axisymmetric disk features (e.g., spiral arms) without a significant increase in random orbital energy.

    \item {\bf Suppressed Disk Heating:}  
    During outward migration, BOB maintains a nearly constant vertical oscillation amplitude, even though stellar populations in FIRE-2 typically experience significant kinematic heating with age.
    This suggests that cold torquing-driven migration may place clusters in dynamically cooler regions of the disk, thereby insulating them from heating by GMCs whose influence declines with galactocentric radius.

    \item {\bf Satellite-Driven Heating and Inward Migration:}  
    Around 12.4 Gyr, BOB experiences a strong perturbation in response to a $\sim$Sgr dSph-mass satellite passing through the disk. 
    This interaction leads to a dramatic increase in vertical amplitude ($|Z_{\mathrm{max}}| \sim2.75$ kpc), due to a localized warping of the disk and coincides with a reversal in radial migration direction, sending the cluster back inward. 
    These changes are consistent with kinematic heating from a satellite-induced disturbance and reinforce the theoretical expectation that satellite interactions increase vertical energy and promote radial redistribution, particularly for low-dispersion systems in the outer disk.  

    \item {\bf Resilience to Disruption:}  
    Despite its dynamical journey --- including cold torquing-driven outward migration, an extended period of vertical stability, and subsequent satellite-induced heating --- BOB remains gravitationally bound across its full $\sim$5 Gyr evolution. 
    This highlights the ability of OCs to endure complex dynamic histories in realistic galactic environments.

    \item {\bf A Close Analog to Be20:}  
 
    BOB roughly matches Be20’s age, metallicity, and vertical excursions while reproducing a plausible orbital pathway that accounts for its present-day $Z$ height and eccentricity. 
    The migration history BOB offers a scenario for how outer disk clusters with large vertical displacements can form and persist in the MW.
\end{enumerate}

This work highlights the power of high-resolution cosmological simulations for linking observed OC populations to their formation and dynamic histories. The remarkable parallels between properties of BOB and Be20 open new avenues for future studies that combine simulation-based archaeology with precise MW data to better understand cluster survival, migration, and vertical structure in disk galaxies. {We encourage future studies to explore the impact of satellite properties, such as mass and inclination angle, on the resulting orbital properties of OCs.}
\begin{acknowledgments}
We thank Adrian Price-Whelan for helpful discussions on disk–satellite interactions, orbital dynamics, and the use of simulated star clusters to interpret origin of outer disk MW OCs. 
We thank Davoud Masoumi for code review and optimization, correcting a minor error in \texttt{sl\_utilities}, and for sharing a coordinate system framework developed during his time in the Loebman Lab, which informed some aspects of our visualization workflow.
{We thank the anonymous referee for their helpful and constructive feedback that meaningfully strengthened the quality of this manuscript.}
AIW, PMF, and JMO acknowledge support from the National Science Foundation through Astronomy and Astrophysics grant AST-2206541 \& AST-2511389.
JRQ acknowledges support from NSF grant AST-2109234 \& AST-2511388.
SRL acknowledges support from NSF grant AST-2109234 \& AST-2511388 and HST grant AR-16624 from STScI.
AW received support from NSF, via CAREER award AST-2045928 and grant AST-2107772.

KJD respectfully acknowledges that the University of Arizona is home to the O’odham and the Yaqui. 
We respect and honor the ancestral caretakers of the land, from time immemorial until now, and into the future.

Simulations presented here were generated via: XSEDE, supported by NSF grant ACI-1548562; Blue Waters, supported by the NSF; Frontera allocations AST21010 and AST20016, supported by the NSF and TACC; and Pleiades, via the NASA HEC program through the NAS Division at Ames Research Center. 
FIRE-2 simulations are publicly available \citep{Weztel2023} at \href{http://flathub.flatironinstitute.org/fire}{flathub.flatironinstitute.org/fire}. 
Additional FIRE simulation data is available at \href{https://fire.northwestern.edu/data}{fire.northwestern.edu/data}. 
A public version of the \textsc{Gizmo} code is available at \href{http://www.tapir.caltech.edu/~phopkins/Site/GIZMO.html}{tapir.caltech.edu/\textasciitilde phopkins/Site/GIZMO.html}.\\
\end{acknowledgments}

\begin{contribution}
AIW led the identification and long-term tracking of BOB.  
JRQ developed visualizations of the satellite–disk interaction.  
MO provided analysis of the asymmetry of the simulated disk with time.
SRL supervised the simulation-based analysis and coordinated the larger project together with PMF.
PMF provided observational expertise on MW open clusters. 
KJD provided guidance on cluster dynamics and helped interpret the orbital evolution.  
FM provided guidance on global kinematic trends at birth and overtime in the simulated system.
JMO contributed observational context for Be20 including orbital integration.  
HRW provided context with moving groups as a consequence of the merger event.
ED aided in the interpretation of simulations \& the understanding of the significance of satellites on dynamics of clusters.
AW provided assistance in the generation and analysis of the FIRE-2 simulations.
HP provided relevant scientific feedback to the manuscript.
BB contributed with code to identify \& track open clusters in the FIRE simulations.
MC provided visualizations of the gas regions surrounding the cluster at the moments of interaction.
\end{contribution}

\bibliography{Bob_Cluster}{}
\bibliographystyle{aasjournalv7}

\end{document}